\documentclass{llncs}
\usepackage[utf8]{inputenc}
\usepackage{graphicx}
\usepackage{tabularx}
\usepackage{array}
\usepackage{pifont}
\usepackage{xspace}
\usepackage{todonotes}
\usepackage{paralist}
\usepackage{subfigure}
\usepackage{thm-restate}
\usepackage{amsmath}

\newcommand{\myparagraph}[1]{\medskip\noindent\textbf{\boldmath #1}}
\newcommand{\cl}{\textsc{ChordLink}\xspace}
\newcommand{\nt}{\textsc{NodeTrix}\xspace}
\newcommand{\nx}{\textrm{n}}
\newcommand{\indeg}{\textrm{in-deg}}

\begin{document}
	
\title{\cl: A New Hybrid Visualization Model\thanks{Work partially supported by: $(i)$ MIUR, under grant 20174LF3T8 ``AHeAD: efficient Algorithms for HArnessing networked Data'', $(ii)$ Dipartimento di Ingegneria - Universit\`a degli Studi di Perugia, under grants RICBASE2017WD and RICBA18WD: ``Algoritmi e sistemi di analisi visuale di reti complesse e di grandi dimensioni''.}
}
\author{Lorenzo Angori\inst{1}, 
Walter Didimo\inst{1}\orcidID{0000-0002-4379-6059},\\
Fabrizio Montecchiani\inst{1}\orcidID{0000-0002-0543-8912}, 
Daniele Pagliuca\inst{1,2},\\
Alessandra Tappini\inst{1}\orcidID{0000-0001-9192-2067}}

\date{}

\institute{ Dipartimento di Ingegneria, Universit\`a degli Studi di Perugia, Italy
\email{\{lorenzo.angori,alessandra.tappini\}@studenti.unipg.it}
\email{\{walter.didimo,fabrizio.montecchiani\}@unipg.it}
\and
	Agenzia delle Entrate, Italy
	\email{daniele.pagliuca@agenziaentrate.it}
}

\maketitle


\begin{abstract}
Many real-world networks are globally sparse but locally dense. Typical examples are social networks, biological networks, and information networks. This double structural nature makes it difficult to adopt a homogeneous visualization model that clearly conveys an overview of the network and the internal structure of its communities at the same time. As a consequence, the use of hybrid visualizations has been proposed. For instance, \nt combines node-link and matrix-based representations (Henry et al., 2007). In this paper we describe \cl, a hybrid visualization model that embeds chord diagrams, used to represent dense subgraphs, into a node-link diagram, which shows the global network structure. The visualization is intuitive and makes it possible to interactively highlight the structure of a community while keeping the rest of the layout stable. We discuss the intriguing algorithmic challenges behind the \cl model, present a prototype system, and illustrate case studies on real-world networks.
\end{abstract}

\section{Introduction}\label{se:introduction}

The challenges in the design of effective visualizations for the analysis of real-world networks are not only related to the size of these networks, but also to the complexity of their structure. In particular, many networks in a variety of application domains are globally sparse but locally dense, i.e., they contain \emph{communities} (or \emph{clusters}) of highly connected nodes, and such communities are loosely connected to each other (see, e.g.,~\cite{fortunato-10,gn-02,pom-09}).  
Typical examples are social networks such as collaboration and financial networks~\cite{DBLP:journals/widm/BediS16,DBLP:journals/vlc/DidimoLM14,okk-04,DBLP:conf/icic/WuHPL10}. Other examples include biological networks (e.g., metabolic and protein-protein interaction networks) and information networks; see, e.g.,~\cite{DBLP:journals/computer/FlakeLGC02,DBLP:journals/bioinformatics/HolmeHJ03,DBLP:conf/cibb/MahmoudMRR13}. A visual exploration of these networks should allow users to perform two main tasks~\cite{DBLP:conf/vl/Shneiderman96}:  \textsf{(T1)} getting an overview of the high-level structure of the network; \textsf{(T2)} identifying and analyzing in detail the communities of the network. However, the heterogeneity of the network connectivity level makes it difficult to adopt a homogeneous visualization that supports both the aforementioned tasks simultaneously.
\\
\indent This scenario naturally motivates the use of \emph{hybrid visualizations} that combine different drawing styles, depending on the connectivity degree of the various portions of the network. A notable example is \nt~\cite{hfm-dhvsn-07}, which adopts a node-link diagram to represent the (sparse) global structure of the network and the more compact matrix representation to visualize denser subgraphs; the user can select the portions of the diagram to be represented as adjacency matrices.

\begin{figure}[t]
	\centering
	\includegraphics[width=0.93\columnwidth]{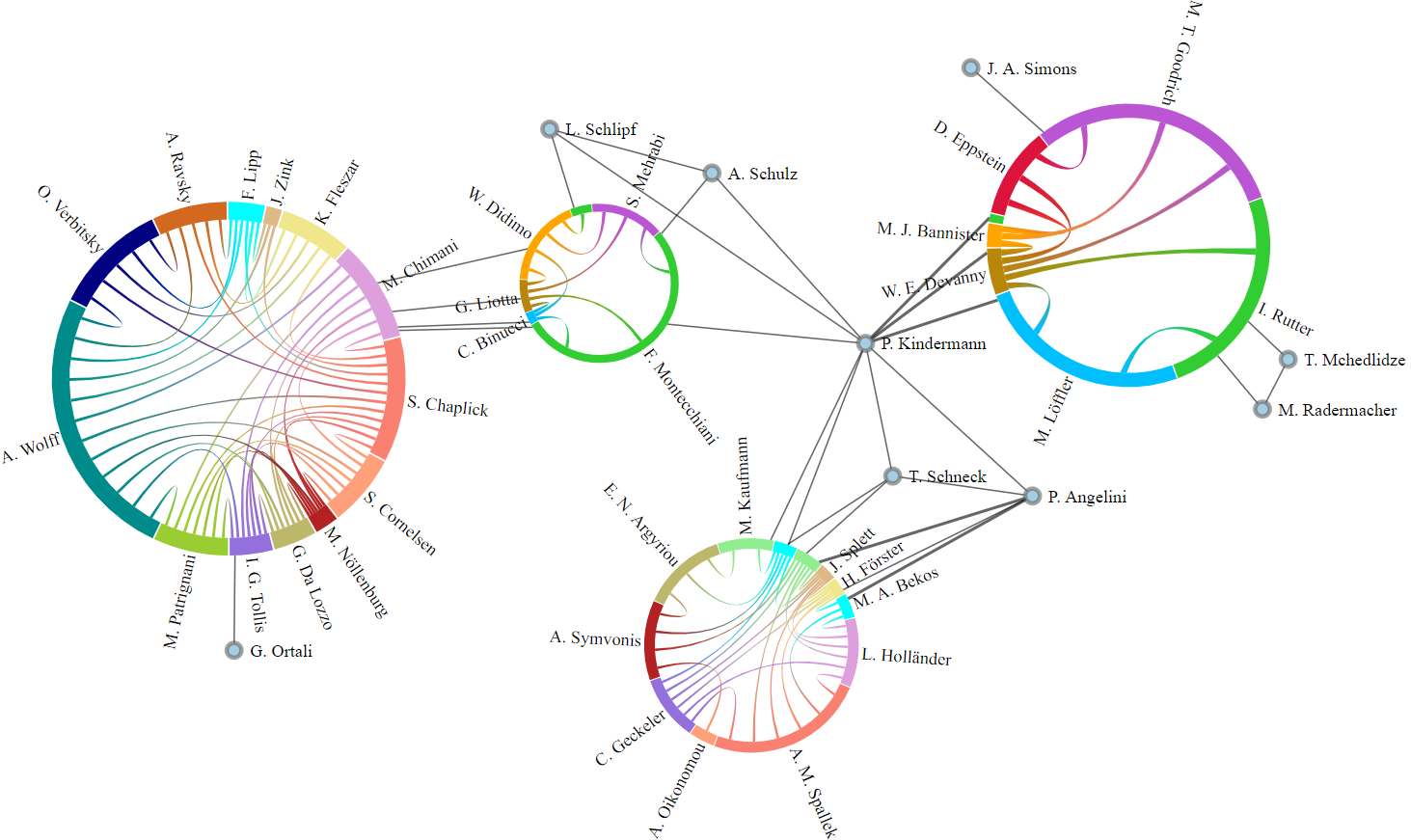}
	\caption{A \cl visualization of a co-authorship network. The drawing has four clusters, represented as chord diagrams. In each chord diagram, circular arcs of the same color are copies of the same author. For example, in the smallest cluster, F.~Montecchiani has two (green) copies, each connected to some nodes external to the cluster.}\label{fi:intro}
\end{figure}
\smallskip
\noindent \textbf{Contribution.} 
Inspired by \nt, we aim to design a hybrid visualization model that supports tasks \textsf{(T1)} and \textsf{(T2)}, and that can be integrated into an interactive visual analytics system. In particular, our design is driven by two main requirements: \textsf{(R1)} the model must support the drawing stability throughout the user interaction,
so to maintain the user's mental map during an interactive analysis of the network;
\textsf{(R2)} the drawing styles to convey the different portions of the network should be intuitive for non-expert users,
as for a node-link representation. Our contribution is as follows:

(i) We propose \cl, a new model that embeds \emph{chord diagrams}, used for the visualization of dense subgraphs (communities), into a node-link diagram, which shows the global network structure (Section~\ref{se:model}). Chord diagrams are an extension of circular drawings, where nodes are represented as circular arcs instead of points (see, e.g.,~\cite{circos}). Figure~\ref{fi:intro} shows a \cl visualization.

(ii) As a proof-of-concept of our model, we describe a prototype system that implements it and we discuss some case studies on different kinds of real-world networks, namely fiscal networks and co-authorship networks (Section~\ref{se:system}). A short video of the system can be found at {\small \url{https://youtu.be/ezphnPEdA8Y}}.

(iii) Finally, our model introduces new optimization problems (Section~\ref{sse:algorithms}) that are of independent interest, and that may inspire future research (Section~\ref{se:conclusions}).

\smallskip
\noindent \textbf{Methodology.}
The \cl model represents a community $C$ selected in a node-link diagram
$\Gamma$ as a specific type of chord diagram, which we denote as $\Gamma(C)$.
Regarding~\textsf{(R1)}, a suitable replication of the nodes of $C$ allows us to preserve the geometry of the nodes and edges outside $\Gamma(C)$; this avoids new edge crossings out of the cluster and supports the user's mental map during an interactive analysis of the network. Such a node-replication also gives additional freedom to reduce the number of edge crossings in $\Gamma(C)$. Regarding~\textsf{(R2)}, the representation $\Gamma(C)$ remains intuitive for users who are familiar with the node-link style, because an edge in $C$ is still represented as a geometric curve. This makes it easy, for example, to recognize paths in $C$, a basic task that is sometimes difficult to perform in a matrix-based representation~\cite{DBLP:journals/ivs/GhoniemFC05,hfm-dhvsn-07}.


\section{Related Work}\label{se:related-work}

Early works in graph visualization propose hybrid models that combine Euler/Venn Diagrams, used to represent inclusion relationships between sets of objects, with Jordan arcs, which convey other types of relationships between these sets~\cite{DBLP:journals/cacm/Harel88,DBLP:conf/vl/SindreGJ93}. Similar drawing styles are extensively used to represent \emph{compound graphs}, where the nodes are hierarchically grouped into clusters and where there can be binary relationships between clusters other than between nodes (see, e.g.,~\cite{DBLP:journals/isci/DogrusozGCCD09,DBLP:conf/dagstuhl/1999dg,DBLP:series/sseke/Sugiyama02} for surveys on the subject). Hybrid visualizations that mix node-link and treemaps are also studied~\cite{fwdap-03,DBLP:conf/infovis/ZhaoMC05}, sometimes in terms of algorithmic techniques for quick computation of clustered layouts~\cite{DBLP:journals/isci/DidimoM14,DBLP:conf/apvis/MuelderM08}. 

The \nt model is the first attempt to visually convey both the global structure of a sparse network and its locally dense subgraphs by combining node-link and matrix-based representations~\cite{hfm-dhvsn-07}. This work has inspired a subsequent array of papers, either devoted to the development of visual analytics systems for complex graphs or focused on the theoretical properties of visualizations in the \nt model. In the first direction, an interesting variant of the \nt model is proposed in~\cite{bbdlpp-valg-11}; while in \nt the clusters represented as an adjacency matrix are selected by the user, in~\cite{bbdlpp-valg-11} the set of clusters is computed by the drawing algorithm so that the resulting graph of clusters (drawn as an orthogonal layout) is planar; the user can choose the drawing style inside each cluster region, including the possibility of using a matrix-based representation. 
In the second direction, several papers study the so-called \emph{hybrid planarity} testing problem, both in the \nt model~\cite{ddfp-cnrcg-jgaa-17,dlprt-ntptsc-19} and in a different model where clusters are intersection graphs of geometric objects~\cite{addfpr-ilrg-17}. This problem asks whether a given graph admits a hybrid visualization such that the edges represented as geometric links do not cross any cluster region and do not cross each other. Also, complexity results on a relaxation of the hybrid planarity testing problem are given in~\cite{DBLP:conf/walcom/GiacomoLLRT19}; similar to \cl, this relaxation allows for a limited replication of the nodes of a cluster, but in~\cite{DBLP:conf/walcom/GiacomoLLRT19} the clusters are defined by the algorithm and intra-cluster edges are not considered.  

Our \cl model uses a specific type of chord diagram to represent clusters. Chord diagrams are effectively adopted in several visualization systems to analyze dense networks in various contexts, including comparative genomics~\cite{circos}, urban mobility trajectories~\cite{Gabrielli2014}, and software profiling on distributed graph processing systems~\cite{DBLP:journals/fgcs/ArleoDLM18}. Other applications of chord diagrams can be found at {\small \url{http://www.circos.ca/}}. They have also been extended to support hierarchical data sets (see, e.g.,~\cite{DBLP:conf/grapp/ArgyriouSV14,DBLP:journals/tvcg/Holten06}). We finally remark that the use of circular layouts for visualizing clustered graphs is proposed in~\cite{DBLP:conf/gd/SixT03}. In that approach, the node set of the input network is partitioned into user-defined clusters, and each cluster is represented as a circular layout with nodes drawn as points and edges drawn as straight segments; hence, each node of the network belongs to a circular layout and the whole drawing of the network is computed by knowing in advance the set of clusters. In the \cl model we assume that the user can define the clusters interactively, and that the drawing of the network must be updated accordingly, while controlling the drawing stability.

\section{The \cl Model}\label{se:model}

Let $G=(V,E)$ be a network and let $\Gamma$ be a node-link diagram of $G$.
The \cl model is conceived to work in an interactive system, in which the user can iteratively select a cluster $C$ of nodes
in $\Gamma$
and the system automatically redraws the subgraph $G[C]$ induced by $C$ as a chord diagram $\Gamma(C)$. The nodes of $C$ are required to lie within a topologically connected region of the plane (e.g., within a circular or a rectangular region); the drawing of nodes and edges of $\Gamma$ out of $G[C]$ should change as little as possible to enforce stability. 

If a node $w \in C$ is connected to a node outside~$C$, we say that $w$ is \emph{extrovert}, else $w$ is \emph{introvert}. To maintain the drawing outside $\Gamma(C)$ stable, the \cl model allows for a suitable replication of the nodes. Namely, every extrovert node $w \in C$ can have multiple occurrences in $\Gamma(C)$, while an introvert node of $C$ will occur exactly once in $\Gamma(C)$. The occurrences of $w$ are called \emph{copies} of~$w$. A copy of $w$ is represented in $\Gamma(C)$ by a circular arc $c_w$, coinciding with a portion of the circumference of $\Gamma(C)$. The set of arcs $c_w$, over all copies of the nodes $w$ of $C$, partitions the circumference of $\Gamma(C)$.   
An edge $(u,w) \notin G[C]$, with $u \notin C$ and $w \in C$, is drawn as a straight-line segment incident to one of the circular arcs $c_w$. An edge $(w,z) \in G[C]$ is drawn as a simple curve, called \emph{chord}, connecting one of the circular arcs $c_w$ to one of the circular arcs~$c_z$. 

\subsection{General Strategy}\label{sse:general-strategy}
Assume that all nodes of a selected cluster $C$ in $\Gamma$ lie in a circular region $R(C)$ and that all the other nodes of $\Gamma$ are outside $R(C)$; also, assume that no node of $C$ is located exactly at the center of $R(C)$ (otherwise slightly perturb the region). According to the \cl model, we locally redraw $\Gamma$ so that the boundary of the chord diagram $\Gamma(C)$ coincides with the boundary of $R(C)$. This is done through a general strategy that consists of the following phases (see Fig.~\ref{fi:general-strategy}):

\begin{figure}[t]
	\centering
	\subfigure[\sf Initial Drawing]{\includegraphics[width=0.48\columnwidth]{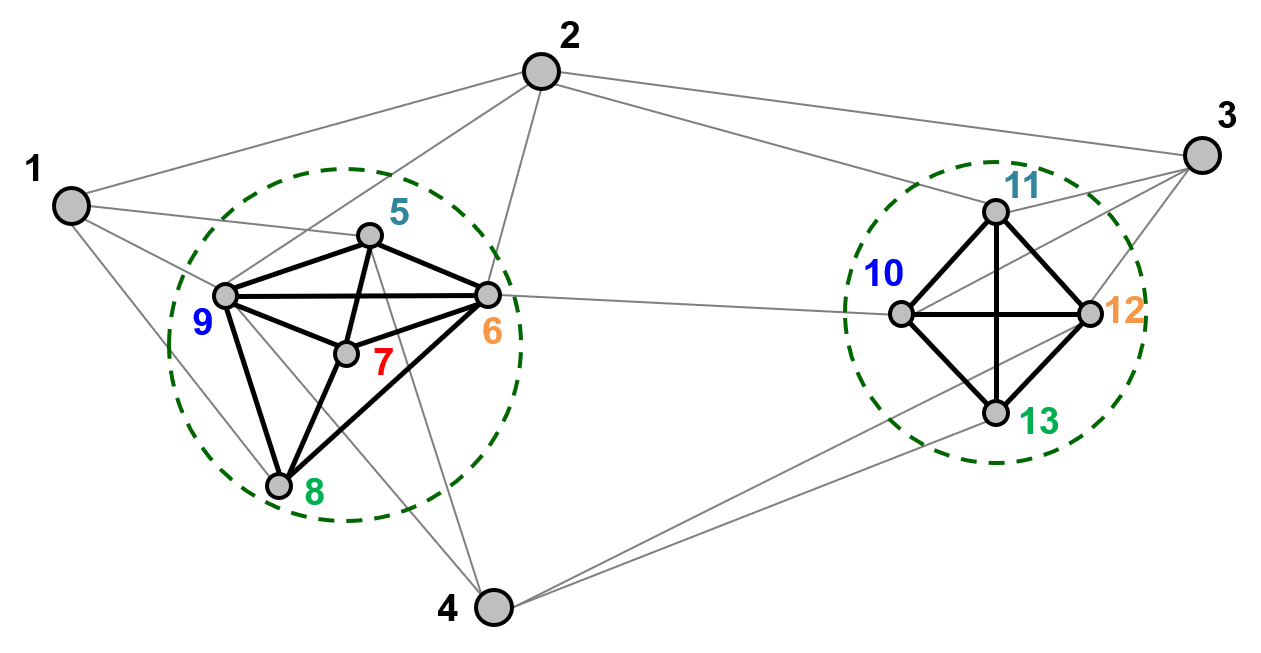}\label{fi:input}}\hfill
	\subfigure[\sf NodeReplication]{\includegraphics[width=0.48\columnwidth]{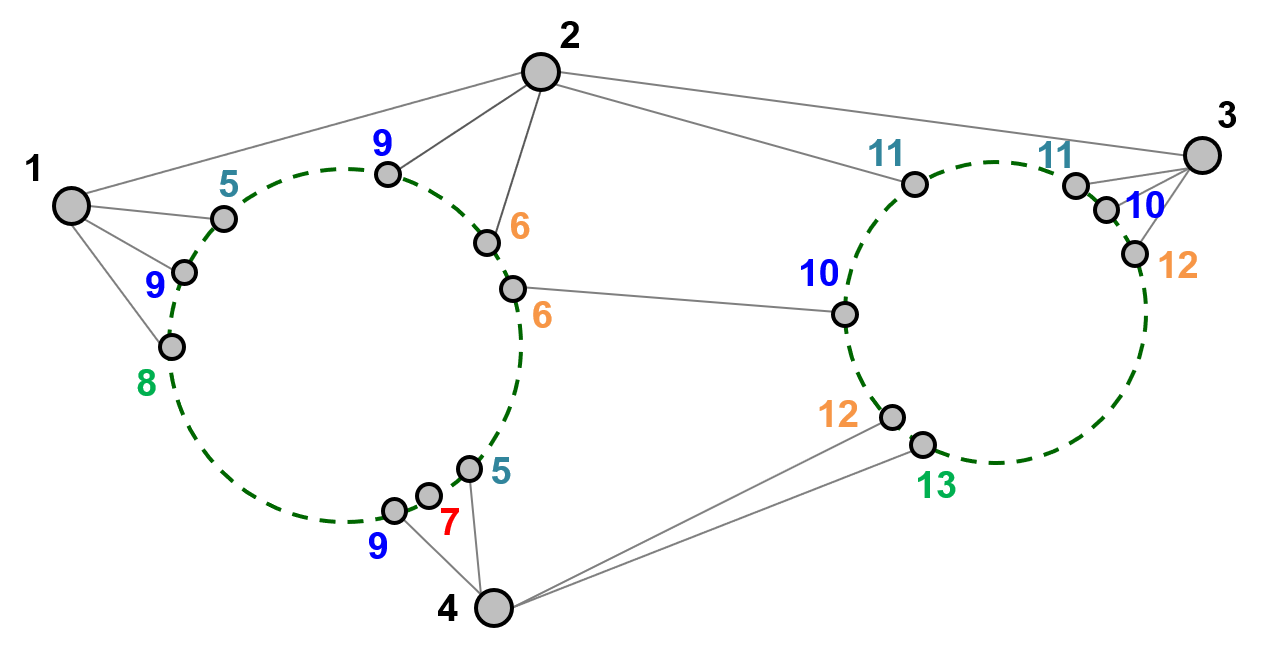}\label{fi:nodereplication}}\hfill
	\subfigure[\sf NodePermutation]{\includegraphics[width=0.48\columnwidth]{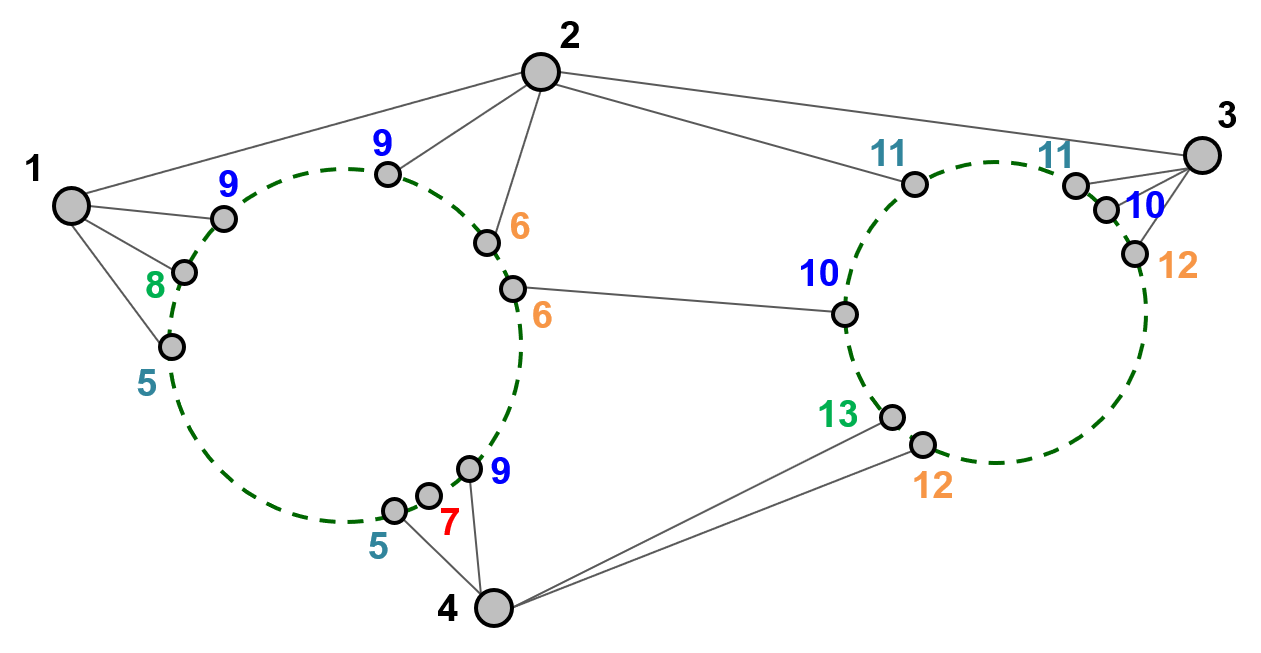}\label{fi:nodepermutation}}\hfill
	\subfigure[\sf NodeMerging+ChordInsertion]{\includegraphics[width=0.48\columnwidth]{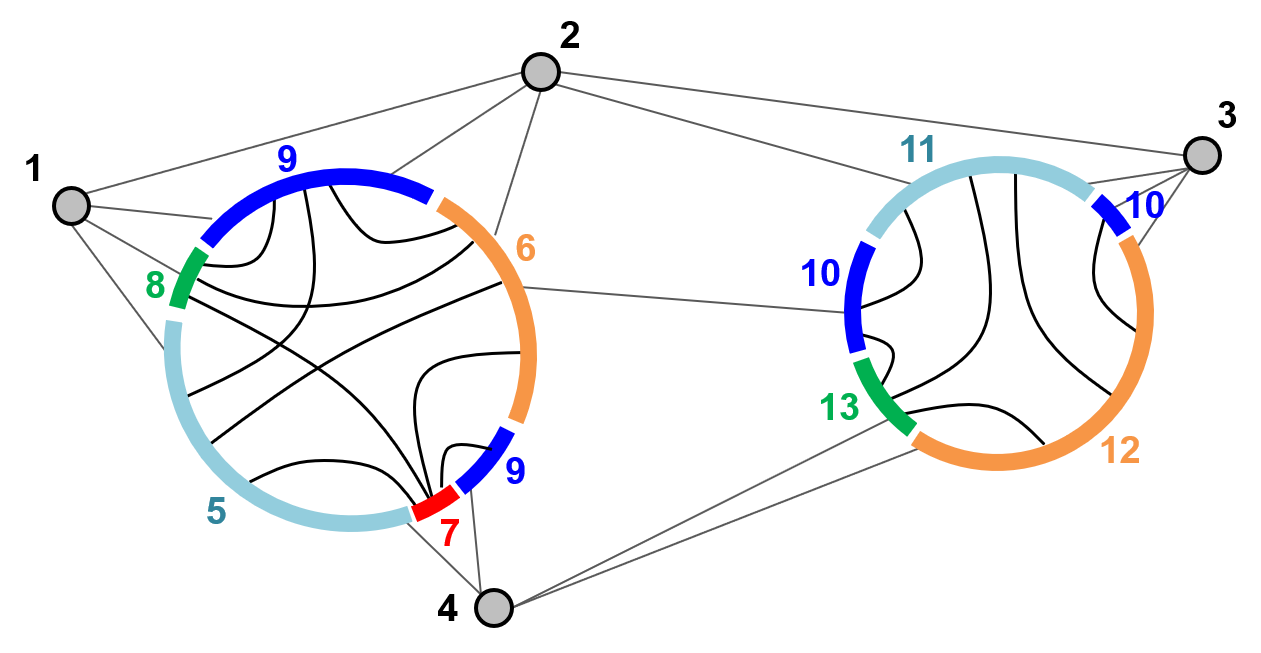}\label{fi:nodemerging-chordinsertion}}
	\caption{Illustration of the general strategy for the \cl model. (a) An initial node-link diagram with two selected clusters (dashed regions). (b) Drawing after the {\sf NodeReplication} phase. (c) Output of the {\sf NodePermutation} phase; for example, in the left cluster the copies of the nodes adjacent to 1 and to 4 are permuted so to reduce the number of non-consecutive copies of 5 and 9. (d) Final drawing after the {\sf NodeMerging} and {\sf ChordInsertion} phases; chords are inserted so to minimize their number of crossings.}\label{fi:general-strategy}
\end{figure}
	\smallskip
	\noindent \textsf{NodeReplication}. For each extrovert node $w \in C$ connected to a node $u \notin C$, create a copy $v$ of $w$ at the intersection point between $(u,w)$ and the boundary of $R(C)$, and replace the segment $\overline{uw}$ with its subsegment $\overline{uv}$. For each introvert node $w \in C$, create a unique copy of $w$ at the intersection point between the boundary of $R(C)$ and the radius of $R(C)$ passing through~$w$. Then, remove all the elements of $\Gamma$ that are properly inside $R(C)$. At the end we have a circular sequence of copies of the nodes of $C$ along the boundary of $R(C)$; two copies of the same node may not be consecutive in this sequence.
	
	\smallskip
	\noindent \textsf{NodePermutation}. Permute the copies of the nodes of $C$ along the boundary of $R(C)$ in such a way to minimize the total number of non-consecutive copies of the same node. To preserve the geometry of the drawing outside $R(C)$, two copies can be permuted only if they are adjacent to the same node $u \notin C$.
	
	\smallskip
	\noindent \textsf{NodeMerging}. For each maximal subsequence of consecutive copies of a node $w$ (possibly a single copy) along the boundary of $R(C)$, replace all these copies by a circular arc $c_w$ that spans at least the whole subsequence.
	
	\smallskip
	\noindent \textsf{ChordInsertion}. For each edge $(w,z) \in G[C]$, select one of the copies $c_w$ and one of the copies $c_z$, and insert a chord inside $R(C)$ connecting $c_w$ and $c_z$. This selection can be done in order to optimize some desired function; for example, one can try to minimize the total number of crossings between chords and/or to maximize the angles formed by two crossing chords.

\subsection{Algorithms}\label{sse:algorithms}

In the following we describe specific algorithms to solve the optimization problems posed by the {\sf NodePermutation} and {\sf ChordInsertion} phases. In Appendix~\ref{se:app-algorithms} we explain how to handle the {\sf NodeMerging} phase and the case in which for a selected cluster there is not a circular region that includes exactly its nodes.

\smallskip
\noindent \textbf{Algorithm for the \textsf{NodePermutation} phase.} 
Let $C$ be a selected cluster in the current drawing $\Gamma$. 
The optimization problem in the {\sf NodePermutation} phase asks to find a permutation of the copies of the nodes of $C$ along the boundary of $R(C)$ such that the total number of non-consecutive copies of the same node is minimized. However, to preserve the geometry of the links outside $R(C)$ (thus avoiding the introduction of edge crossings), two copies can be permuted only if they have a common neighbor $u \notin C$. Formally, we model the problem as follows.

Let $u_1, u_2, \dots, u_k$ be the set of nodes not in $C$ that are adjacent to some node of $C$. For each $u_i$ $(i=1, \dots, k)$, denote by $\langle v_{i,1}, v_{i,2}, \dots, v_{i,h_i} \rangle$ the clockwise sequence of copies of extrovert nodes of $C$ along $R(C)$ attached to $u_i$. For example, assume that $C$ is the left-side cluster in Fig.~\ref{fi:nodereplication}; if we set $u_1=1$, $u_2=2$, $u_3=10$, and $u_4=4$  then we have: $\langle v_{1,1}=8, v_{1,2}=9, v_{1,3}=5 \rangle$; $\langle v_{2,1}=9, v_{2,2}=6 \rangle$; $\langle v_{3,1}=6 \rangle$; $\langle v_{4,1}=5, v_{4,2}=9 \rangle$. The sequence $\langle v_{i,1}, v_{i,2}, \dots, v_{i,h_i} \rangle$ is called the \emph{group of $u_i$}. Clearly, two elements of the same group never represent copies of the same node of $C$. 
Denote by $\cal E$ the set of copies of the extrovert nodes of $C$ on the boundary of $R(C)$. Suppose that $v \in \cal E$ is a copy of a node $w \in C$ and that $\nx(v)$ is the next copy of $w$ encountered by walking clockwise on the boundary of $R(C)$. We denote by $\chi(v,\nx(v))$ the \emph{cost of} $\{v,\nx(v)\}$ and we define it as follows: $\chi(v,\nx(v))=0$ if no copies of nodes of $C$ are encountered between $v$ and $\nx(v)$ while walking clockwise on the boundary of $R(C)$; $\chi(v,\nx(v))=1$ otherwise. Our optimization problem asks to find a permutation of the copies in the group of $u_i$ (for each $i=1, \dots, k$) that minimizes the objective function $\sum_{v\in \cal E} \chi(v,\nx(v))$.  

We describe a dynamic programming algorithm that we designed with the aim of computing
an exact solution for this optimization problem when all the copies in each group are consecutive along the boundary of $R(C)$ (like in Fig.~\ref{fi:general-strategy}); if this is not the case, our algorithm is used as a heuristic for the problem.  
If all the copies of each group are consecutive, two node permutations $\pi$ and $\pi'$ yield the same cost if for each group the first element is the same in both $\pi$ and $\pi'$ and the same holds for the last element. Hence, it suffices to minimize the pairs of consecutive groups such that their two neighboring elements are copies of different nodes. More formally, let $B_0, B_1, \dots, B_{k-1}$ be the clockwise sequence of groups along $R(C)$, starting from an arbitrary group $B_0$. For each group $B_i$,  let $f_i$ and $l_i$ be its first and its last element, respectively, i.e., $l_i$ and $f_{i+1}$ (indexes taken modulo $k$) are consecutive along $R(C)$. Our dynamic programming formulation considers the cost of choosing the first and the last element of $B_i$ assuming that this choice has been already done for the groups $B_{i+1},\dots,B_{k-1}$. Namely, denote by $O_i(v_{i,j},v_{i,z})$ the cost of choosing $f_i=v_{i,j}$ and $l_i=v_{i,z}$. For each possible pairs of elements $v_{i,j},v_{i,z}$ in $B_i$ and $v_{i+1,j'},v_{i+1,z'}$ in $B_{i+1}$, the following holds:

\begin{equation}
	O_i(v_{i,j},v_{i,z})= O_{i+1}(v_{i+1,j'},v_{i+1,z'}) +
	\begin{cases}
		0, & \mbox{if } v_{{i+1},j'}=v_{i,z} \\
		1, & \mbox{if } v_{{i+1},j'}\neq v_{i,z} 
	\end{cases}
\end{equation}
The optimal solution is then $\chi_{opt} = \min_{v_{0,j},v_{0,z} \in B_0}{O_0(v_{0,j},v_{0,z})}$. 
To solve the above recurrence we fix $f_0$ and compute a table of size $\sum_{i=0}^{k-1} {h_i \choose 2} \le m^2$, where $m$ is the number of edges of $G$. We repeat this procedure for each of the $h_0 \le m$ possible values of $f_0$ and we select the optimal solution among them; this algorithm takes $O(m^3)$ time. Note that, to speed up the algorithm, the elements $v_{i,j}$ such that there is no element $v_{i+1,j'}=v_{i,j}$ in $B_{i+1}$ (resp. $v_{i-1,j'}=v_{i,j}$ in $B_{i-1}$) can  be ignored, since selecting them as first or last element of $B_i$ always increases the cost of the solution. In particular, we first remove them in a preprocessing step, and then reinsert them in any position between~$f_i$ and~$l_i$. 

\begin{figure}[t]
	\centering
	\subfigure[]{\includegraphics[width=0.25\columnwidth]{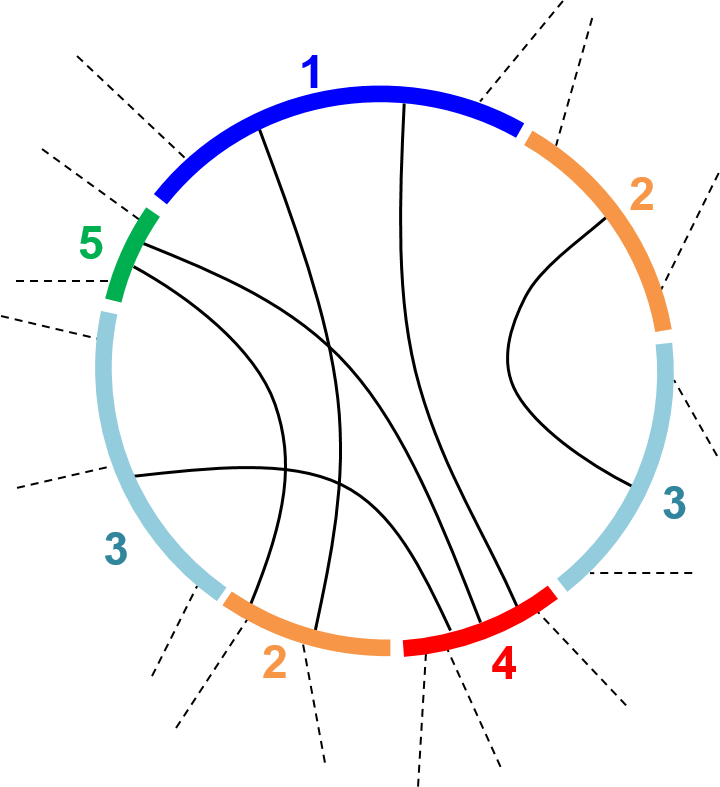}\label{fi:chordinsertion-ex-a}}
	\hfil
	\subfigure[]{\includegraphics[width=0.25\columnwidth]{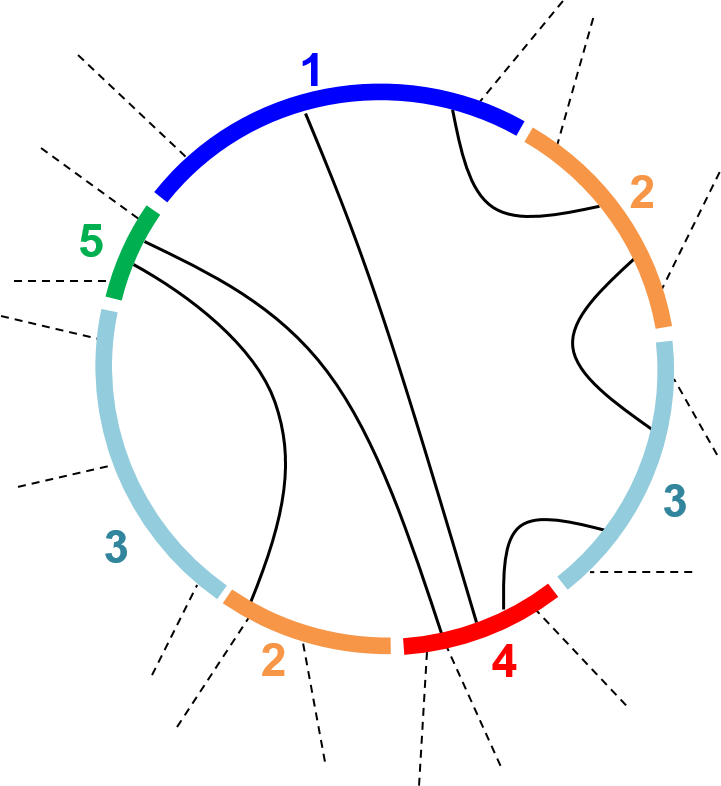}\label{fi:chordinsertion-ex-b}}
	\caption{Example of different choices in the \textsf{ChordInsertion} phase. The set of chords in each drawing represents the edges $(1,2)$, $(1,4)$, $(2,3)$, $(2,5)$, $(3,4)$, $(4,5)$. In (a) the chords form $3$ crossings, while in (b) they do not cross, due to a more convenient choice of the representative pair of arcs for the edges $(1,2)$ and $(3,4)$. The dashed lines represent stubs of possible outside edges incident to the cluster.}\label{fi:chordinsertion-ex}
\end{figure}

\smallskip
\noindent \textbf{Algorithm for the \textsf{ChordInsertion} phase.} 
In this phase, for each edge $(w,z) \in G[C]$ we have to select one of the circular arcs $c_w$ associated with $w$ and one of the circular arcs $c_z$ associated with $z$, and we add a chord connecting $c_w$ to $c_z$. The specific selection of a pair $\{c_w,c_z\}$ for each edge $(w,z)$
determines the total number of crossings between chords. For example, Fig.~\ref{fi:chordinsertion-ex} shows a schematic illustration of two different chord diagrams for a cluster~$C$. The cluster has seven circular arcs, associated with nodes $1$, $2$, $3$, $4$, $5$; the edges of $G[C]$ are $(1,2)$, $(1,4)$, $(2,3)$, $(2,5)$, $(3,4)$, and $(4,5)$. The chords representing these edges cause in total $3$ crossings in Fig.~\ref{fi:chordinsertion-ex-a}, while they do not cross in the drawing of Fig.~\ref{fi:chordinsertion-ex-b}, where we have chosen a different pair of arcs for the edges $(1,2)$ and $(3,4)$.   
\\
\indent Our algorithm for selecting the set of chords aims to minimize the number of crossings and to maximize the minimum angle at a crossing point of two crossing chords. This optimization goal is motivated by several works that show the negative impact of the number of crossings (e.g.,~\cite{DBLP:journals/iwc/Purchase00,DBLP:journals/ese/PurchaseCA02,DBLP:journals/ivs/WarePCM02}) and in particular of small crossing angles (e.g.,~\cite{DBLP:journals/vlc/HuangEH14,DBLP:journals/jgaa/HuangHE07}) in graph layouts.
\\
\indent We model the above optimization problem as follows. We assume that each circular arc $c_w$ is collapsed into a single point $p_w$, coinciding with the center of $c_w$. Once the set of chords incident to $p_w$ is decided by the algorithm, we expand back $p_w$ to $c_w$ and equally distribute the chords incident to $p_w$ along $c_w$. Note that, the number of crossings between non-adjacent chords only depends on the circular order of their end-points along $R(C)$ and not on their exact position. Hence, two non-adjacent chords $(p_w,p_z)$, $(p_x, p_y)$ cross if and only the corresponding chords $(c_w,c_z)$, $(c_x, c_y)$ cross, independent of the position of the end-points of the chords along $c_w$, $c_z$, $c_x$, and $c_y$. Also, two adjacent chords $(p_w,p_z)$ and $(p_w,p_x)$ never cross, and therefore the corresponding chords $(c_w,c_z)$ and $(c_w,c_x)$ will not cross if we use the same circular order. Moreover, if $(p_w,p_z)$ and $(p_x,p_y)$ are two crossing chords, we denote by $a(\overline{wz},\overline{xy})$ the minimum angle formed by the segments $\overline{wz}$ and $\overline{xy}$ at their crossing point; this gives an estimation of the crossing angular resolution of the two chords if each chord is drawn as a monotone curve approximating the straight segment between its end-points.
For any two chords $e_{wz} = (p_w,p_z)$ and $e_{xy} = (p_x,p_y)$, we define the \emph{cost of} the unordered pair $\{e_{wz}, e_{xy}\}$ as a function $\alpha(e_{wz},e_{xy})$ such that: 
$\alpha(e_{wz},e_{xy})=0$ if $e_{wz}$ and $e_{xy}$ do not cross; $\alpha(e_{wz},e_{xy})=1-a(\overline{wz},\overline{xy})/\pi$ otherwise. Since $a(\overline{wz},\overline{xy}) \in (0,\pi/2]$, we have $\alpha(e_{wz},e_{xy}) \in [0.5,1)$. We aim to select a set $S$ of chords for the edges of $G[C]$ that minimizes the cost function $\alpha(S) = \sum_{\{e_{wz},e_{xy}\} \in S \times S}\alpha(e_{wz},e_{xy})$.
\\
\indent To solve this problem we use a heuristic algorithm based on a greedy strategy. Let $E(C)$ be the set of edges of $G[C]$ and let $E_1(C) \subseteq E(C)$ be the subset of edges having one representative chord $(p_w, p_z)$, i.e., $(w,z) \in E_1(C)$ if and only if $w$ and $z$ have a unique copy on the boundary of $R(C)$. Also, let $E_2(C) = E(C) \setminus E_1(C)$ be the remaining subset of edges of $G[C]$. For example, in the cluster $C$ of Fig.~\ref{fi:chordinsertion-ex} we have $E_1(C)=\{(4,5)\}$ and $E_2(C)=\{(1,2), (1,4), (2,3), (2,5), (3,4)\}$. Our algorithm first adds to the drawing $\Gamma(C)$ the chords representing the edges of $E_1(C)$ (in any order), because for these edges there are no alternative choices. After that, the algorithm executes $|E_2(C)|$ iterations. Each iteration $i$ ($1 \leq i \leq |E_2(C)|$) removes an edge $(w,z)$ from $E_2(C)$ and adds to the drawing one of its representative chords $(p_w,p_z)$. More precisely, let $S_0$ be the set of chords added for the edges in $E_1(C)$ and let $S_i$ denote the set of chords added at the end of iteration $i$.   
At the beginning of iteration $i$, for each edge $(w,z) \in E_2(C)$ and for each chord $(p_w,p_z)$ that is representative of $(w,z)$,  the algorithm computes the cost of inserting $(p_w,p_z)$ in the current drawing, i.e., the cost $\alpha(S_{i-1} \cup \{(p_w,p_z)\})$; then it selects the chord that yields the minimum cost and removes from $E_2(C)$ the corresponding edge. Denote by $S'$ the whole set of representative chords for the edges of $E(C)$. Since the cost $\alpha(S_{i-1} \cup \{(p_w,p_z)\})$ can be easily computed in $O(|S_{i-1}|)$ time from the cost $\alpha(S_{i-1})$ and from the set of chords in $S_{i-1}$, and since $|S_{i-1}| = O(|E(C)|)$, the whole greedy algorithm takes $O(|S'||E(C)|^2)$ time.

\section{A Prototype System}\label{se:system}

As a proof-of-concept of the \cl model, we realized a prototype system that implements it.
The system is developed in Javascript (so to run in a Web browser) and the implementation uses the \textsc{D3.js} library~\cite{d3}, {\small \url{https://d3js.org}}. We first describe the main features of the system interface and its interaction functionalities. Then, we discuss two case studies that show how the system can be used to perform the analysis tasks {\sf (T1)} and {\sf (T2)} on different kinds of real networks, namely a fiscal network and a co-authorship network.       

\noindent \textbf{Interface and Interaction.}
Through the interface of our system, the user can import a network in the GML file format~\cite{gml}.
The system initially computes a node-link diagram of the network using a force-directed algorithm; we exploit an implementation available in the \textsc{D3.js} library. The interface supports the visualization of weighted edges by using different levels of edge thickness to convey this information.
The user can execute some common operations, like node movement, zooming, and panning. Node labels can be displayed according to different policies. One can show/hide all labels at the same time or enable/disable each label individually. Alternatively, the system can automatically manage the visualization of labels based on node-degrees and on the current zoom level of the layout (labels of low-degree nodes are hidden after a zoom-out operation). Regardless of the labeling policy, a mouse-hover operation on a node or on an edge causes the display of a tooltip that reports the label of that element.

In order to represent a desired cluster $C$ as a chord diagram $\Gamma(C)$, the user can select the nodes of $C$ in the layout (e.g., through a rectangular region selection). The visualization of $\Gamma(C)$ is such that: $(i)$ All the circular arcs $c_w$ associated with the same node $w \in C$ are assigned the same color; the label of $w$ is displayed near to one of its corresponding arcs, namely the longest one. $(ii)$ Each chord between two arcs $c_w$ and $c_z$ has a color that gradually goes from the color of $c_w$ to that of $c_z$; this helps to visually detect the end-nodes of the chord. $(iii)$ The size of each chord reflects the weight of the corresponding edge (the maximum thickness for the chords in $\Gamma(C)$ depends on the minimum length of the circular arcs and on their inner degree). A mouse-hover operation on a circular arc $c_w$ of $\Gamma(C)$ highlights all the arcs associated with $w$, as well as all the edges incident to $c_w$ (see Fig.~\ref{fi:mouse-hover-arc} in Appendix~\ref{se:app-system}).
The user can move a chord diagram $\Gamma(C)$ or drag a node $u \notin C$ to drop it in $\Gamma(C)$; this operation adds $u$ to $C$ and causes an immediate update of the drawing. The user can click on $\Gamma(C)$ to collapse it into a single \emph{cluster-node} (whose size is proportional to the number of nodes in $C$); a click operation on a cluster-node expands back it into the original chord diagram. Collapsing/expanding each cluster individually helps focusing on specific portions of the network without losing the general context where they are embedded (see Fig.~\ref{fi:collapsed} in Appendix~\ref{se:app-system}).

\begin{figure}[tbp]
	\centering
	\includegraphics[width=.91\columnwidth]{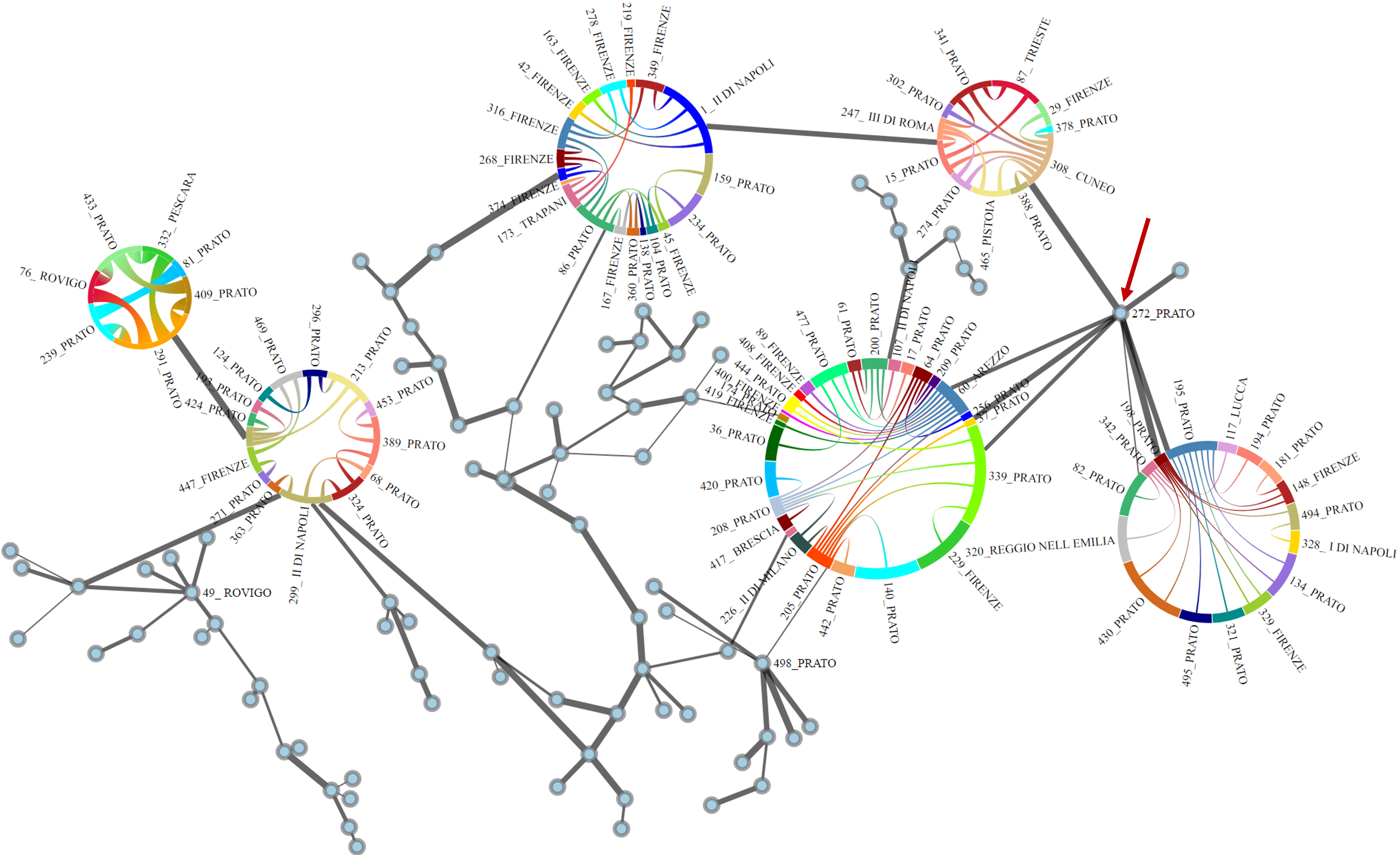}
	\caption{A visualization obtained by selecting some communities in a node-link diagram.}\label{fi:cs-1-2}
\end{figure}

%
\noindent \textbf{Case Studies: Fiscal Networks.}
The first case study falls into the domain of fiscal risk analysis. We considered a real network of taxpayers and their economic transactions. The network is provided by the IRV (Italian Revenue Agency) and refers to a portion of data for the fiscal year 2014, consisting of 174 subjects with high fiscal risk and 200 economic transactions between them~\cite{DBLP:journals/dss/DidimoGLMP18}. 
Figure~\ref{fi:cs-1-2} depicts a \cl visualization of this network computed by our system after the selection of six clusters (Fig.~\ref{fi:cs-1-1} in Appendix~\ref{se:app-system} reports the initial node-link diagram).
The thickness of an edge $(u,v)$ reflects the amount of transactions between $u$ and $v$ in the considered year (we discretized the range of amounts into 5 values of thickness). For privacy reasons data are anonymized; a node's label reports the ID number and the geographic area of the corresponding taxpayer.
\\ \indent Regarding task {\sf (T1)}, we observe that the network consists of several communities and of few nodes with high degree.
A visual analysis of the network reveals that the node with ID 272 (marked with an arrow in the figure) acts as a broker between three communities, since it has strong connections with them.
Regarding task {\sf (T2)}, the chord diagram of each community makes it possible to analyze the connections between its nodes, by overcoming the node overlaps in the node-link diagram. The position of nodes and the geometry of edges outside the chord diagrams do not change with respect to the initial node-link diagram, since all nodes of every selected community lie in a circular region not containing other nodes of the network. Focusing on the rightmost chord diagram $\Gamma(C)$ in Fig.~\ref{fi:cs-1-2}, we can see that the node with ID~272 is connected to two nodes of high degree inside $\Gamma(C)$ (those with IDs~195 and~198), which belong to the same geographic area.
An analyst of the IRV identified this subgraph as a suspicious scheme characterized by several economic transactions, where the seller is a so-called ``missing trader'' with serious tax irregularities (omitted VAT payments or tax declarations); nodes with IDs~195 and~198 are missing traders. From a deepest inspection of the connections in $\Gamma(C)$ and from additional attributes of its taxpayers, the analyst confirmed the presence of a tax evasion pattern.
Similar conclusions were derived from the analysis of other communities in the network.

\begin{figure}[tbp]
	\centering
	\includegraphics[width=.95\columnwidth]{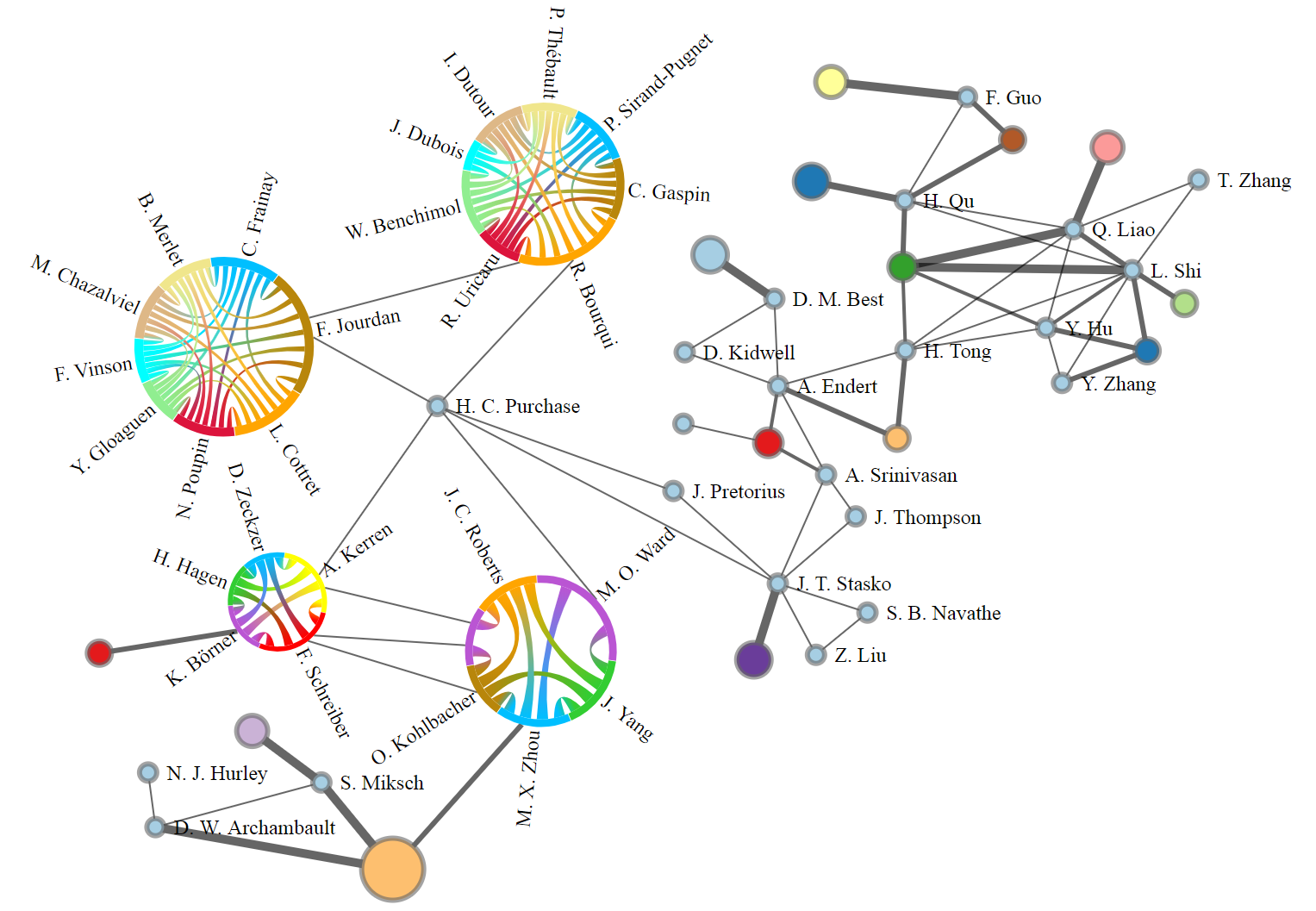}
	\caption{A co-authorship network extracted from DBLP. Bigger nodes are cluster-nodes.}\label{fi:cs-2-1}
\end{figure}

\smallskip
\noindent \textbf{Case Studies: Co-authorship Networks.}
The second case study considers co-authorship networks extracted from the DBLP dataset~\cite{dblp}, which contains publication data in computer science. Through a query consisting of keywords and Boolean operators, one can retrieve a set of publications on a desired topic. We use the results returned by DBLP to construct networks where nodes are authors and edges indicate co-authorships, weighted by the number of papers shared by their end-nodes. Nodes are labeled with authors' names and edges with the titles of the corresponding publications. 
\\
\indent We performed the query \textsf{``network \textsf{AND} visualization''} and limited to 500 the number of search results (i.e., publications) to be returned. The resulting network consists of 1766 nodes, 3780 edges, and 382 connected components. The largest of these components contains 118 nodes and 322 edges. A \cl visualization of this component is shown in Fig.~\ref{fi:cs-2-1}, where several dense portions of the original node-link layout have been identified as communities. To make the diagram easier to read, some communities (on the left side) have been expanded and some others (on the right side) have been collapsed. We now discuss some findings that involve tasks~{\sf (T1)} and~{\sf (T2)} in an interleaved manner. 
%
\\
\indent From the general structure of the clustered network one can clearly distinguish several central actors. For example, on the left side of the drawing we can observe that H.~C.~Purchase is connected to four distinct communities. Following the links incident to this author and the connections between the related authors inside the clusters, we can see that H.~C.~Purchase forms a $3$-cycle with A.~Kerren and M.~O.~Ward (this author has two copies in his cluster), who fall into two distinct communities. By exploring the edge labels, we see that this cycle originates from a work titled \textsf{``Introduction to Multivariate Network Visualization''}, while the communities to which A.~Kerren and M.~O.~Ward belong mainly derive from the works \textsf{``Heterogeneous Networks on Multiple Levels''} and \textsf{``Novel Visual Metaphors for Multivariate Networks''}, respectively.
By analyzing the literature more in detail, one can observe that these three works appear in the same book, referring to the Dagstuhl Seminar \textsf{Multivariate Network Visualization}.
The orange cluster-node in the bottom of the drawing, call~it~$C$, seems to be strongly related to nodes S.~Miksch, D.~W.~Archambault, and M.~X.~Zhou. Indeed, the links of these three authors with $C$ refer to a common work, \textsf{``Temporal Multivariate Networks''}. Since D.~W.~Archambault has only two connections with nodes outside $C$, it seems reasonable to move it inside $C$ by a drag operation.
\\
\indent If we analyze this community in detail (Fig.~\ref{fi:cs-2-2} in Appendix~\ref{se:app-system} shows its chord diagram), the connections reveal that the aforementioned work has other $5$ authors in addition to the $3$ already cited. Two of them, K.~Ma and C.~Muelder, have a connection thicker than the other pairs of nodes, which indicates a stronger cooperation. Also, there are two nodes of $C$, namely S.~Diehl and F.~Tzeng, that are loosely connected in this cluster. We deduce that it would be convenient to keep them out of the community, even if the original node-link diagram locates them very close to the other nodes of~$C$.  
%

\section{Final Remarks and Future Work}\label{se:conclusions}
The \cl model proposed in this paper is a new kind of hybrid visualization.
It can complement previous models conceived for the visual analysis of networks that are globally sparse but locally dense.
Among its advantages, \cl makes it possible to keep the visualization stable during the interaction.
This is especially true when the nodes of a community, that is going to be represented as a chord diagram, are close to each other in the node-link layout (which is most often the case if it is computed by a force-directed algorithm). Nonetheless, \cl has also some clear limits. In particular, the readability of a chord diagram may degrade when the size of a cluster increases; our current visualization can be effectively used for clusters up to 20-25 nodes, while it becomes less effective for bigger clusters.

Besides these considerations, we believe that the \cl model opens the way for intriguing research directions:
(i) We conjecture that the optimization problems at the core of a \cl visualization are computationally hard. It would be interesting to prove NP-hardness and to design new algorithms to be compared with our heuristics.
(ii) It may be worth developing a system that combines the \cl and the \nt models, allowing users to switch from a visualization to the other for each cluster. This would merge the advantages of both models.
(iii) One can exploit an automatic clustering algorithm for the \cl model, e.g., one that guarantees the planarity of the inter-cluster graph~\cite{bbdlpp-valg-11}. 

\newpage
\bibliographystyle{splncs04}
\bibliography{biblio}

\begin{thebibliography}{10}
\providecommand{\url}[1]{\texttt{#1}}
\providecommand{\urlprefix}{URL }
\providecommand{\doi}[1]{https://doi.org/#1}

\bibitem{addfpr-ilrg-17}
Angelini, P., {Da Lozzo}, G., {Di Battista}, G., Frati, F., Patrignani, M.,
  Rutter, I.: Intersection-link representations of graphs. Journal of Graph
  Algorithms and Applications  \textbf{21}(4),  731--755 (2017).
  \doi{10.7155/jgaa.00437}

\bibitem{DBLP:conf/grapp/ArgyriouSV14}
Argyriou, E.N., Symvonis, A., Vassiliou, V.: A fraud detection visualization
  system utilizing radial drawings and heat-maps. In: Laramee, R.S., Kerren,
  A., Braz, J. (eds.) {IVAPP} 2014. pp. 153--160. SciTePress (2014).
  \doi{10.5220/0004735501530160}

\bibitem{DBLP:journals/fgcs/ArleoDLM18}
Arleo, A., Didimo, W., Liotta, G., Montecchiani, F.: Profiling distributed
  graph processing systems through visual analytics. Future Generation Comp.
  Syst.  \textbf{87},  43--57 (2018). \doi{10.1016/j.future.2018.04.067}

\bibitem{bbdlpp-valg-11}
Batagelj, V., Brandenburg, F., Didimo, W., Liotta, G., Palladino, P.,
  Patrignani, M.: Visual analysis of large graphs using ({X},{Y})-{Clustering}
  and hybrid visualizations. {IEEE} Trans. Vis. Comput. Graph.
  \textbf{17}(11),  1587--1598 (2011). \doi{10.1109/TVCG.2010.265}

\bibitem{DBLP:journals/widm/BediS16}
Bedi, P., Sharma, C.: Community detection in social networks. Wiley
  Interdiscip. Rev. Data Min. Knowl. Discov.  \textbf{6}(3),  115--135 (2016).
  \doi{10.1002/widm.1178}

\bibitem{d3}
Bostock, M., Ogievetsky, V., Heer, J.: D{\({^3}\)} {Data-Driven Documents}.
  {IEEE} Trans. Vis. Comput. Graph.  \textbf{17}(12),  2301--2309 (2011).
  \doi{10.1109/TVCG.2011.185}

\bibitem{ddfp-cnrcg-jgaa-17}
{Da Lozzo}, G., {Di Battista}, G., Frati, F., Patrignani, M.: Computing
  {NodeTrix} representations of clustered graphs. Journal of Graph Algorithms
  and Applications  \textbf{22}(2),  139--176 (2018). \doi{10.7155/jgaa.00461}

\bibitem{DBLP:conf/walcom/GiacomoLLRT19}
{Di Giacomo}, E., Lenhart, W.J., Liotta, G., Randolph, T.W., Tappini, A.: (k,
  p)-planarity: {A} relaxation of hybrid planarity. In: {WALCOM}. Lecture Notes
  in Computer Science, vol. 11355, pp. 148--159. Springer (2019)

\bibitem{dlprt-ntptsc-19}
Di~Giacomo, E., Liotta, G., Patrignani, M., Rutter, I., Tappini, A.: {NodeTrix}
  planarity testing with small clusters. Algorithmica  (May 2019).
  \doi{10.1007/s00453-019-00585-6}

\bibitem{DBLP:journals/dss/DidimoGLMP18}
Didimo, W., Giamminonni, L., Liotta, G., Montecchiani, F., Pagliuca, D.: A
  visual analytics system to support tax evasion discovery. Decision Support
  Systems  \textbf{110},  71--83 (2018). \doi{10.1016/j.dss.2018.03.008}

\bibitem{DBLP:journals/vlc/DidimoLM14}
Didimo, W., Liotta, G., Montecchiani, F.: Network visualization for financial
  crime detection. J. Vis. Lang. Comput.  \textbf{25}(4),  433--451 (2014).
  \doi{10.1016/j.jvlc.2014.01.002}

\bibitem{DBLP:journals/isci/DidimoM14}
Didimo, W., Montecchiani, F.: Fast layout computation of clustered networks:
  Algorithmic advances and experimental analysis. Inf. Sci.  \textbf{260},
  185--199 (2014). \doi{10.1016/j.ins.2013.09.048}

\bibitem{DBLP:journals/isci/DogrusozGCCD09}
Dogrus{\"{o}}z, U., Giral, E., Cetintas, A., Civril, A., Demir, E.: A layout
  algorithm for undirected compound graphs. Inf. Sci.  \textbf{179}(7),
  980--994 (2009). \doi{10.1016/j.ins.2008.11.017}

\bibitem{fwdap-03}
Fekete, J.D., Wang, D., Dang, N., Aris, A., Plaisant, C. (eds.): Overlaying
  graph links on treemaps. IEEE Symposium on Information Visualization
  Conference Compendium (demonstration) (2003)

\bibitem{DBLP:journals/computer/FlakeLGC02}
Flake, G.W., Lawrence, S., Giles, C.L., Coetzee, F.: Self-organization and
  identification of web communities. {IEEE} Computer  \textbf{35}(3),  66--71
  (2002). \doi{10.1109/2.989932}

\bibitem{fortunato-10}
Fortunato, S.: Community detection in graphs. Physics Reports
  \textbf{486}(3-5),  75--174 (2010). \doi{10.1016/j.physrep.2009.11.002}

\bibitem{Gabrielli2014}
Gabrielli, L., Rinzivillo, S., Ronzano, F., Villatoro, D.: From tweets to
  semantic trajectories: Mining anomalous urban mobility patterns. In: Nin, J.,
  Villatoro, D. (eds.) {CitiSens} 2013. pp. 26--35. Springer (2014).
  \doi{10.1007/978-3-319-04178-0\_3}

\bibitem{DBLP:journals/ivs/GhoniemFC05}
Ghoniem, M., Fekete, J., Castagliola, P.: On the readability of graphs using
  node-link and matrix-based representations: a controlled experiment and
  statistical analysis. Information Visualization  \textbf{4}(2),  114--135
  (2005)

\bibitem{gn-02}
Girvan, M., Newman, M.E.J.: Community structure in social and biological
  networks. Proc. Natl. Acad. Sci. USA  \textbf{99}(12),  7821--7826 (2002).
  \doi{10.1073/pnas.122653799}

\bibitem{DBLP:journals/cacm/Harel88}
Harel, D.: On visual formalisms. Commun. {ACM}  \textbf{31}(5),  514--530
  (1988). \doi{10.1145/42411.42414}

\bibitem{hfm-dhvsn-07}
Henry, N., Fekete, J., McGuffin, M.J.: {NodeTrix}: {A} hybrid visualization of
  social networks. {IEEE} Trans. Vis. Comput. Graph.  \textbf{13}(6),
  1302--1309 (2007). \doi{10.1109/TVCG.2007.70582}

\bibitem{gml}
Himsolt, M.: {GML}: A portable graph file format (technical report
  {U}niversität {P}assau) (2010)

\bibitem{DBLP:journals/bioinformatics/HolmeHJ03}
Holme, P., Huss, M., Jeong, H.: Subnetwork hierarchies of biochemical pathways.
  Bioinformatics  \textbf{19}(4),  532--538 (2003).
  \doi{10.1093/bioinformatics/btg033}

\bibitem{DBLP:journals/tvcg/Holten06}
Holten, D.: Hierarchical edge bundles: Visualization of adjacency relations in
  hierarchical data. {IEEE} Trans. Vis. Comput. Graph.  \textbf{12}(5),
  741--748 (2006). \doi{10.1109/TVCG.2006.147}

\bibitem{DBLP:journals/vlc/HuangEH14}
Huang, W., Eades, P., Hong, S.: Larger crossing angles make graphs easier to
  read. J. Vis. Lang. Comput.  \textbf{25}(4),  452--465 (2014).
  \doi{10.1016/j.jvlc.2014.03.001}

\bibitem{DBLP:journals/jgaa/HuangHE07}
Huang, W., Hong, S., Eades, P.: Effects of sociogram drawing conventions and
  edge crossings in social network visualization. J. Graph Algorithms Appl.
  \textbf{11}(2),  397--429 (2007). \doi{10.7155/jgaa.00152}

\bibitem{DBLP:conf/dagstuhl/1999dg}
Kaufmann, M., Wagner, D. (eds.): Drawing Graphs, Methods and Models (the book
  grow out of a Dagstuhl Seminar, April 1999), Lecture Notes in Computer
  Science, vol.~2025. Springer (2001). \doi{10.1007/3-540-44969-8}

\bibitem{circos}
Krzywinski, M., Schein, J., Birol, n., Connors, J., Gascoyne, R., Horsman, D.,
  Jones, S.J., Marra, M.A.: Circos: An information aesthetic for comparative
  genomics. Genome Res.  \textbf{19}(9),  1639--1645 (2009).
  \doi{10.1101/gr.092759.109}

\bibitem{dblp}
Ley, M.: The {DBLP} computer science bibliography,
  \url{https://dblp.uni-trier.de}

\bibitem{DBLP:conf/cibb/MahmoudMRR13}
Mahmoud, H., Masulli, F., Rovetta, S., Russo, G.: Community detection in
  protein-protein interaction networks using spectral and graph approaches. In:
  {CIBB}. Lecture Notes in Computer Science, vol.~8452, pp. 62--75. Springer
  (2013). \doi{10.1007/978-3-319-09042-9\_5}

\bibitem{DBLP:conf/apvis/MuelderM08}
Muelder, C., Ma, K.: A treemap based method for rapid layout of large graphs.
  In: PacificVis. pp. 231--238. {IEEE} Computer Society (2008).
  \doi{10.1109/PACIFICVIS.2008.4475481}

\bibitem{okk-04}
Onnela, J., Kaski, K., Kert\'esz, J.: Clustering and information in correlation
  based financial networks. The European Physical Journal B-Condensed Matter
  and Complex Systems  \textbf{38}(2),  353--362 (2004).
  \doi{10.1140/epjb/e2004-00128-7}

\bibitem{pom-09}
Porter, M.A., Onnela, J.P., Mucha, P.J.: Communities in networks. Notices of
  the American Mathematical Society  \textbf{56},  1082--1097, 1164--1166
  (2009)

\bibitem{DBLP:journals/iwc/Purchase00}
Purchase, H.C.: Effective information visualisation: {A} study of graph drawing
  aesthetics and algorithms. Interacting with Computers  \textbf{13}(2),
  147--162 (2000). \doi{10.1016/S0953-5438(00)00032-1}

\bibitem{DBLP:journals/ese/PurchaseCA02}
Purchase, H.C., Carrington, D.A., Allder, J.: Empirical evaluation of
  aesthetics-based graph layout. Empirical Software Engineering  \textbf{7}(3),
   233--255 (2002)

\bibitem{DBLP:conf/vl/Shneiderman96}
Shneiderman, B.: The eyes have it: {A} task by data type taxonomy for
  information visualizations. In: Proceedings of the 1996 {IEEE} Symposium on
  Visual Languages, Boulder, Colorado, USA, September 3-6, 1996. pp. 336--343
  (1996). \doi{10.1109/VL.1996.545307}

\bibitem{DBLP:conf/vl/SindreGJ93}
Sindre, G., Gulla, B., Jokstad, H.G.: Onion graphs: {A}sthetics and layout. In:
  {VL}. pp. 287--291. {IEEE} Computer Society (1993).
  \doi{10.1109/VL.1993.269613}

\bibitem{DBLP:conf/gd/SixT03}
Six, J.M., Tollis, I.G.: A framework for user-grouped circular drawings. In:
  Graph Drawing. Lecture Notes in Computer Science, vol.~2912, pp. 135--146.
  Springer (2003). \doi{10.1007/978-3-540-24595-7\_13}

\bibitem{DBLP:series/sseke/Sugiyama02}
Sugiyama, K.: Graph Drawing and Applications for Software and Knowledge
  Engineers, Series on Software Engineering and Knowledge Engineering, vol.~11.
  WorldScientific (2002). \doi{10.1142/4902}

\bibitem{DBLP:journals/ivs/WarePCM02}
Ware, C., Purchase, H.C., Colpoys, L., McGill, M.: Cognitive measurements of
  graph aesthetics. Information Visualization  \textbf{1}(2),  103--110 (2002).
  \doi{10.1057/palgrave.ivs.9500013}

\bibitem{DBLP:conf/icic/WuHPL10}
Wu, H., He, J., Pei, Y., Long, X.: Finding research community in collaboration
  network with expertise profiling. In: {ICIC} {(1)}. Lecture Notes in Computer
  Science, vol.~6215, pp. 337--344. Springer (2010).
  \doi{10.1007/978-3-642-14922-1\_42}

\bibitem{DBLP:conf/infovis/ZhaoMC05}
Zhao, S., McGuffin, M.J., Chignell, M.H.: Elastic hierarchies: Combining
  treemaps and node-link diagrams. In: {INFOVIS}. pp. 57--64. {IEEE} Computer
  Society (2005). \doi{10.1109/INFVIS.2005.1532129}

\end{thebibliography}

\newpage
\appendix

\section*{Appendix}\label{ap:appendix}

\section{Additional Material for Section~\ref{sse:algorithms}}\label{se:app-algorithms}

\myparagraph{Algorithm for the \textsf{NodeMerging} phase.} In this phase we have to replace each maximal subsequence $S_w$ of copies of the same node $w$ along the boundary of $R(C)$ with a circular arc $c_w$. As already mentioned, $c_w$ must span at least $S_w$, so to keep the incidences of the external edges on $C$ correct. However, within this constraint we can decide to further balance the lengths of each $c_w$ in order to better accommodate the internal edges incident to $c_w$. Denote by $v_w^{start}$ and $v_w^{end}$ the starting and the ending elements in $S_w$ in clockwise order and let $\indeg(c_w)$ be the number of edges of $G[C]$ incident to $c_w$. Initially the extremes of $c_w$ coincide with the positions of $v_w^{start}$ and $v_w^{end}$. Then, for each pair of consecutive arcs $c_w$ and $c_z$ along the boundary of $R(C)$ we move $v_w^{end}$ clockwise and $v_z^{start}$ counterclockwise, until they meet in a point between their original position. The choice of this point is done in such a way that the final length of each $c_w$ is proportional to $\indeg(c_w)$, under the constrains given by the external edges. The fact that $v_w^{end}$ is moved clockwise and $v_z^{start}$ counterclockwise guarantees that the constraint imposed by the external edges is not violated. Finally, to make $v_w^{end}$ and $v_z^{start}$ clearly distinguishable, we create a small gap between them in the drawing, and we guarantee a minimum length for each~$c_w$.     

\myparagraph{Handling non-circular selections.} So far we have assumed that for a cluster $C$ there exists a circular region $R(C)$ that includes all the nodes of $C$ and that excludes all the other nodes of $G$. This is always the case if the user selects a group of nodes by highlighting a circular region. However, if the user is allowed to select a cluster by highlighting a rectangular region or by performing a ``lasso'' selection (i.e., a ``free form'' selection), it might happen that any circular region $R(C)$ that includes all the nodes of the cluster also contains some other nodes. In this case, we locally deform the drawing so that the nodes of $G \setminus G[C]$ are moved outside $R(C)$. Namely, we apply the following strategy. The center $c_0$ of $R(C)$ is set as the barycenter of the nodes of $C$ and the radius of $R(C)$ is set as the minimum radius necessary to include all the nodes of $C$. If $R(C)$ contains some nodes that do not belong to $C$, we translate every node $u \notin C$ radially along the line through $u$ and $c_0$, by a length that: $(i)$ suffices to bring $u$ outside $R(C)$; $(ii)$ decreases for increasing distances of $u$ from $c_0$; $(iii)$ the radial order of the nodes of the drawing with respect to $c_0$ does not change.

\section{Additional Material for Section~\ref{se:system}}\label{se:app-system}

\begin{figure}[h!]
	\centering
	\subfigure[]{\includegraphics[width=0.36\columnwidth]{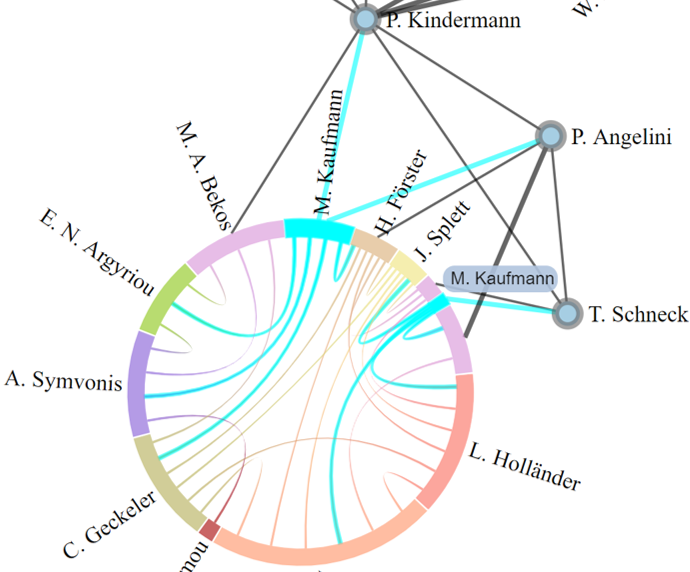}\label{fi:mouse-hover-arc}}
	\hfill
	\subfigure[]{\includegraphics[width=0.63\columnwidth]{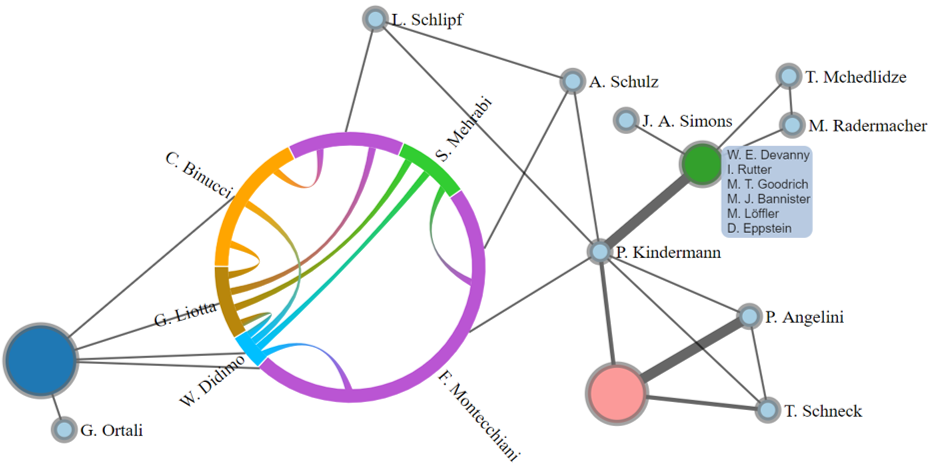}\label{fi:collapsed}}
	\caption{(a) A mouse-hover operation on the circular arc corresponding to ``M. Kaufmann''. (b) A \cl representation where some clusters are collapsed; a mouse-hover on a collapsed cluster opens a tooltip that lists all the authors in the cluster.}\label{fi:interaction}
\end{figure}

\begin{figure}[h!]
	\centering
	\includegraphics[width=1\columnwidth]{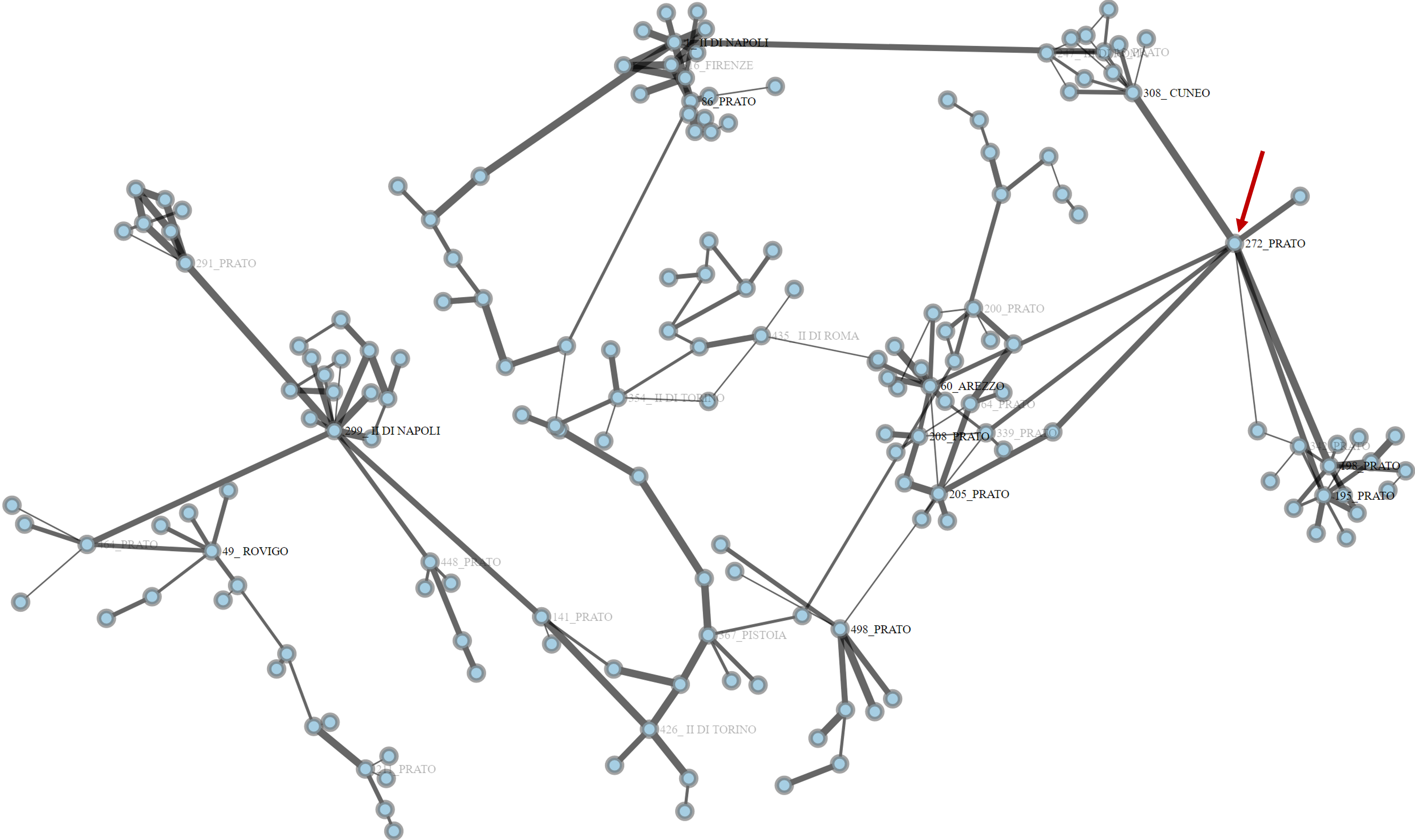}
	\caption{An initial node-link diagram of a fiscal network with 174 nodes and 200 edges.}\label{fi:cs-1-1}
\end{figure}

\begin{figure}[h!]
	\centering
	\includegraphics[width=0.9\columnwidth]{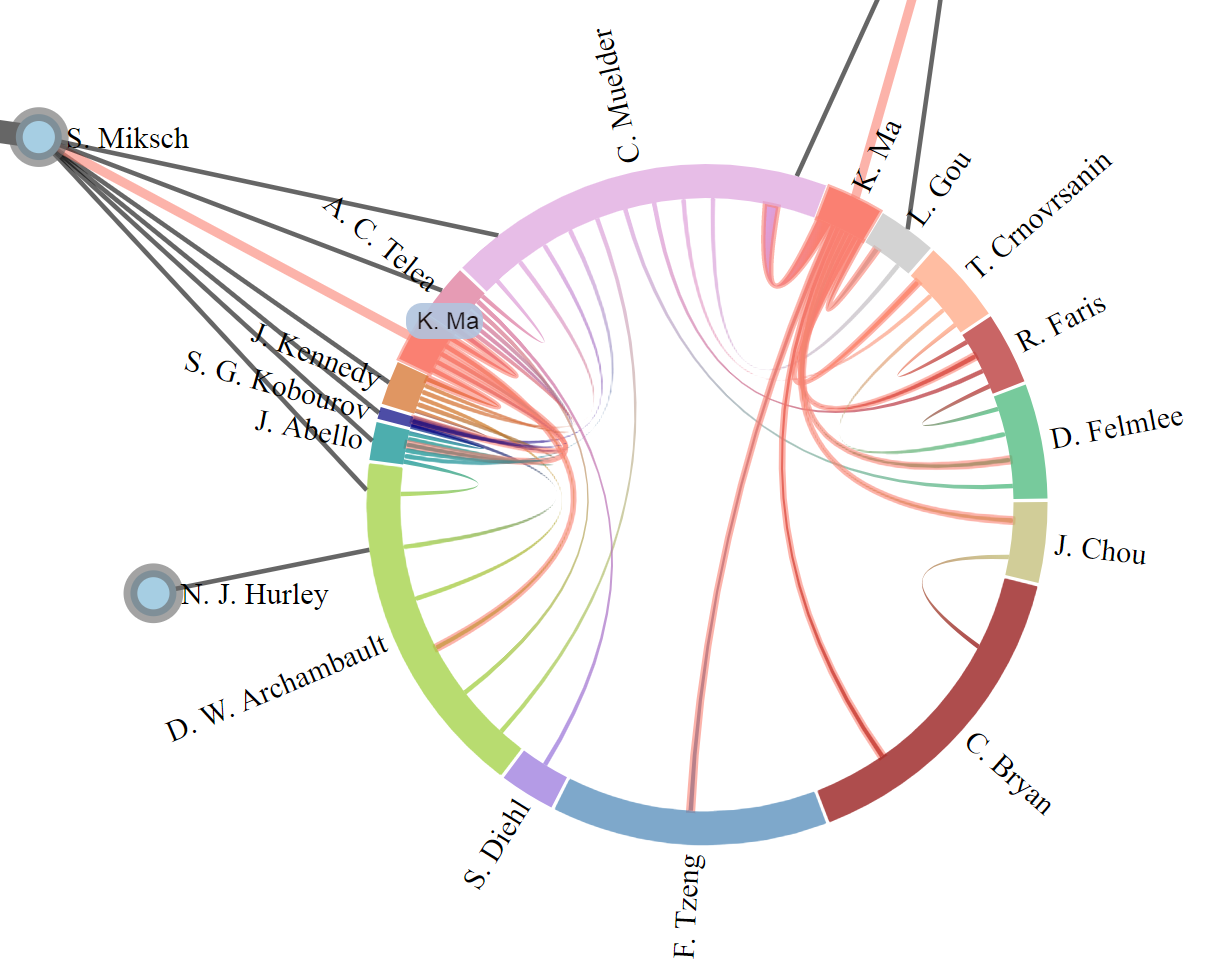}
	\caption{Detailed view of a cluster in the network of Fig.~\ref{fi:cs-2-1}. The mouse is positioned over K.~Ma, thus all the occurrences and the connections of this node are highlighted.}\label{fi:cs-2-2}
\end{figure}

\end{document}